\documentclass[%
 aip,
 amsmath,amssymb,
 reprint,%
]{revtex4-1}

\usepackage{graphicx}
\usepackage{dcolumn}
\usepackage{bm}

\usepackage[utf8]{inputenc}
\usepackage[T1]{fontenc}
\usepackage{mathptmx}
\usepackage{etoolbox}
\usepackage{xcolor}
\usepackage{soul}

\makeatletter
\def\@email#1#2{%
 \endgroup
 \patchcmd{\titleblock@produce}
  {\frontmatter@RRAPformat}
  {\frontmatter@RRAPformat{\produce@RRAP{*#1\href{mailto:#2}{#2}}}\frontmatter@RRAPformat}
  {}{}
}%
\makeatother
\begin{document}

\preprint{AIP/123-QED}

\title[Cosmic-ray tomography of shipping containers]{Cosmic-ray tomography of shipping containers:\\A 
combination of complementary secondary particle and muon 
information using simulations}
\author{Maximilian P\'{e}rez Prada*}%
 \email{m.perezprada@dlr.de}
\author{\'{A}ngel Bueno Rodr\'{i}guez}%
\author{Maurice Stephan}%
\author{Sarah Barnes}%
\affiliation{\textsuperscript{\text{\emph{1)}}} German Aerospace Center (DLR) \\~~~ Institute for the Protection of Maritime Infrastructures \\~~~ Fischkai 1, 27572 Bremerhaven, Germany} 

\date{\today}

\begin{abstract}
Cosmic-ray tomography usually relies on measuring the scattering or transmission of muons produced within cosmic-ray air showers to reconstruct an examined volume of interest (VOI). During the traversing of a VOI, all air shower particles, including muons, interact with the matter within the VOI producing so-called secondary particles. The characteristics of the production of these particles contain additional information about the properties of the examined objects and their materials. However, this approach has not been fully realized practically. Hence, this work aims to study a novel technique to scan shipping containers by comparing and combining the complementary results from stand-alone secondary particles and muon scattering using simulated simplified scenes with a 1 m\textsuperscript{3} cube made out of five different materials located inside the container. The proposed approach for a statistical combination is based on a multi-step procedure centered around a clustering and segmentation algorithm. This ensures a consistent evaluation and comparison of the results before and after the combination focusing on dedicated properties of the reconstructed object. The findings of this work show a potential improvement over the results obtained solely through muon scattering due to the utilization of secondary particle information by applying this novel dual-channel cosmic-ray tomography analysis.
\end{abstract}

\maketitle

\section{\label{sec:Intro}Introduction}
Cosmic-ray tomography (CRT) has become a novel technique for non-destructive imaging among a wide range of applications in the recent past, like monitoring volcanoes, analyzing the internal structure of pyramids and detecting contraband within shipping containers~\cite{bib01,bib02,bib03,bib04,bib05,bib06,bib07,bib08}. CRT utilizes muons from so-called cosmic-ray air showers. These particle showers are produced due to the interaction of high-energy cosmic-ray particles with the nuclei of atoms in earth’s atmosphere and reach down to earth’s surface. At sea level, air showers are mostly composed out of photons, neutrinos, muons, electrons, protons, neutrons and pions~\cite{bib09,bib10,bib11,bib12}. The muons produced in cosmic-ray air showers show high potential for imaging purposes as a natural source of high-energy radiation due to their ability to penetrate deeply into any kind of material. 

Within the field of CRT, either the method of Muon Scattering Tomography (MST) or the Muon Radiography (MR) approach are mainly in use. For applications based on MST, the Coulomb scattering of muons inside the examined volume of interest (VOI) is utilized for image reconstruction as the scattering angle depends on the specific material properties, notably the atomic number and density. One can therefore reconstruct the examined objects and their characteristics by measuring the deflection of the muon trajectory within the VOI~\cite{bib13,bib14}. The method of MR utilizes the muon absorption rate for the image reconstruction, as the number of stopped muons inside a material depends mostly on its density~\cite{bib15,bib16}. 

During the interaction of air shower particles with matter (e.g. through bremsstrahlung or muon capture), additional, so-called secondary particles are produced within the VOI. The production rate and kinematics of secondary particles, mainly photons, electrons and neutrons, are material dependent~\cite{bib17,bib18,bib19,bib20,bib21,bib22,bib23,bib24,bib25}. By measuring these particles, a complementary source of information can be utilized to improve the reconstruction of the examined objects and their properties in comparison to measurements obtained solely from MST or MR. Theoretical and experimental studies on the usage of this complementary set of information up to now mainly focused on the coincidence measurement of secondary particles and incoming air shower muons~\cite{bib26,bib27,bib28,bib29,bib30}. The feasibility of a stand-alone technique has been shown in previous work by the authors in the context of shipping container scans relying entirely on simulation studies. The studies included the analysis of the container content based solely on the information obtained through the measurement of secondary particles under optimal and realistic detection conditions~\cite{bib31,bib32}. Additionally, necessary geometric corrections on the reconstructed objects were investigated, derived and validated~\cite{bib33}.

This work focuses on the first direct comparison and statistical combination of the results acquired through MST and secondary particle analysis (SPA) building upon the methods presented in the previous work by the authors~\cite{bib31,bib32,bib33}, which are the first of its kind for a stand-alone secondary analysis approach. At first, the setup of the simulation including the geometry of the detector and the container is briefly explained. Next, the separate reconstruction methodologies for SPA and MST are presented, which allow for the reconstruction and discrimination of the objects located inside the VOI. Following this, the combination procedure based on a clustering algorithm including the normalization and convolution of the reconstructed objects, as well as its segmentation over the noise-induced background is explained. In addition, the performance metrics used to evaluate the results before and after the combination are described. In conclusion, the results of this first-of-its-kind approach based on a consistent statistical combination of the complementary MST and SPA measurements are presented enabling further studies on such dual-channel analyses.

\section{\label{sec:Sim}Simulation setup}
The simulations are performed with the \textsc{Geant4} Monte Carlo simulation toolkit~\cite{bib34,bib35} to model all interactions between particles and matter. The precomputed and parametrized look-up tables of the Cosmic-ray Shower Library (CRY)~\cite{bib12} are used to reproduce the cosmic-ray air showers and the kinematics of the resulting particles at sea level. The air showers are set to contain muons, photons, electrons, protons, neutrons and pions, as well as a maximum number of 30 particles within an air shower. In addition, the side length of the generation plane is 10~m and the latitude is set to 42 degrees north. For each simulated scene, 100~million air showers are generated, which is equivalent to a scanning time of around 30~minutes, which is realistic under regular operation conditions for a secondary custom control. 

\begin{figure}[h!]
	\centering
	\includegraphics[width=8.6cm]{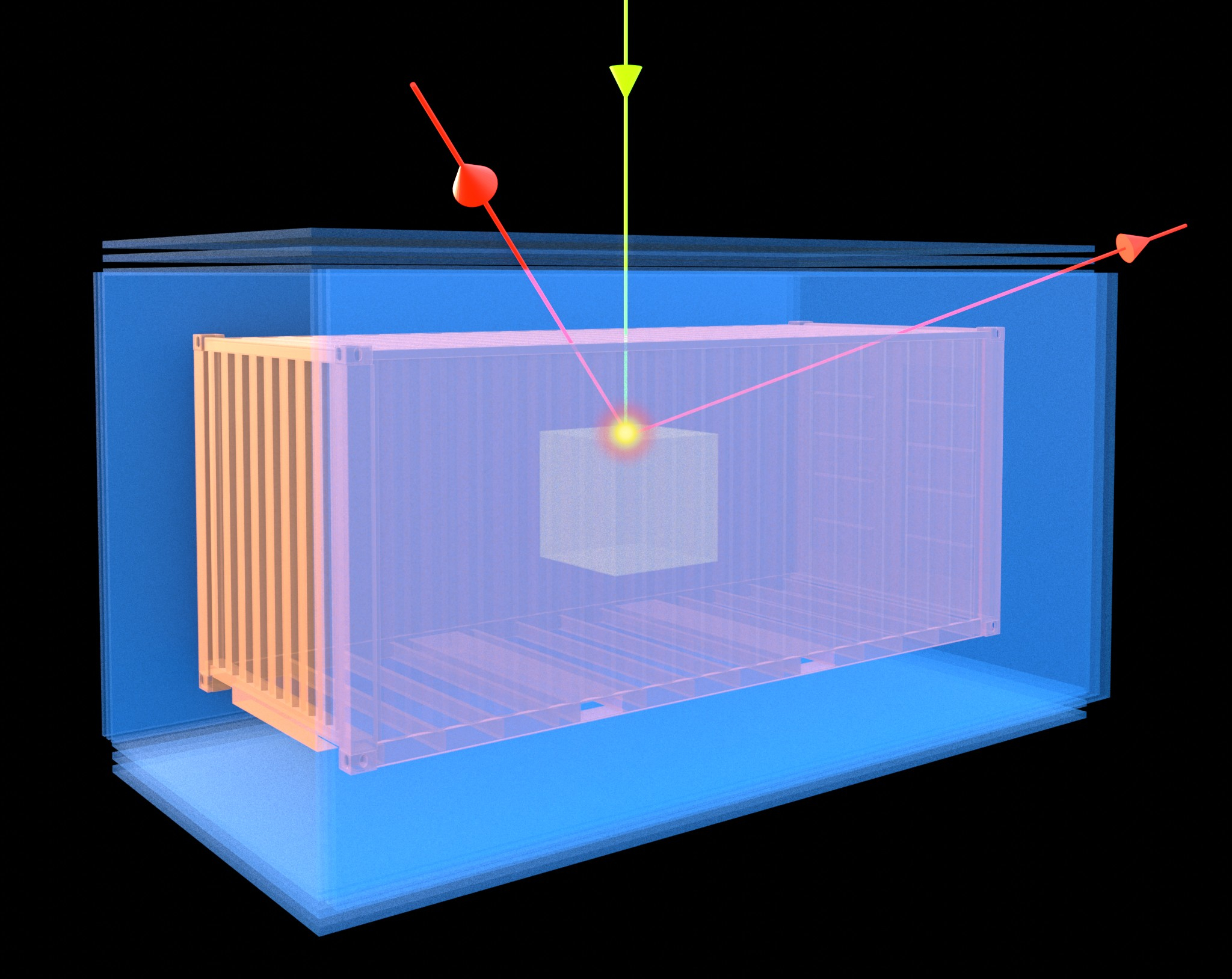}
	\caption{Visualization of the simulation setup. The detector layers are indicated by the blue planes surrounding the container, in which the target object is located. The trajectory of an air shower particle is represented by the green line, while the pink lines show the paths of secondary particles produced during the depicted interaction of the air shower particle with the target cube.}
	\label{figure:Geometry}
\end{figure}

The target object consists of a 1~m\textsuperscript{3} cube, which is located in the middle of a simplified model of a shipping container made out of stainless steel. The size and shape of the target object was chosen to imitate standard cargo objects inside a container, such as intermediate bulk containers or loaded europallets. Five different scenes are generated, each with a different cube material: water, aluminum, vanadium, copper and lead. The origin of the coordinate system is set to be at the center of the container, with the x-axis along the width, the y-axis along the length and the z-axis along the height of the container. The detector configuration is based on a common design used in CRT~\cite{bib08,bib14,bib26,bib29} and consists of four sets of three 50~mm thick polyvinyl toluene based plastic scintillator layers with 10~cm spacing. As depicted in Figure~\ref{figure:Geometry}, two of these sets are arranged vertically, one at the top and one at the bottom, while the other two are positioned on the left and right sides of the container. This configuration delineates a volume of interest with dimensions of 3.5~m~\texttimes~7.0~m~\texttimes~3.5~m. 

Particle information (e.g. energy, particle type, momentum direction) from each detector layer is directly retrieved from \textsc{Geant4}, as neither the particle identification, nor the readout from the scintillator material is emulated. The goal of this work is the development and evaluation of the theoretical groundwork for a first stand-alone secondary particle analysis. Therefore, all studies are conducted on a generalized level without a dedicated detector design in place. The main reason of the placement of the plastic scintillator material is the emulation of the impact of a sensitive detection material on the secondary particle kinematics. 

Particles which don't originate within the VOI or fail to intersect all three detection layers in any of the four detector sets are excluded from the analysis. The estimated trajectory of each particle (muon, photon, neutron and electron) is based on a least squares fitting method using the three detection positions across the three scintillator layers. For simplicity, perfect spatial resolution and a detection efficiency of 100~\% are assumed as these values strongly correlate with the precise detector configuration, which is beyond the scope of this work. Further studies quantifying the impact of realistic detector conditions on the results of the SPA can be found in a previous work by the authors~\cite{bib32}.

\section{\label{section:Reco}Reconstruction methodology}
\subsection{\label{sec:Reco_SPA}Secondary particle analysis (SPA)}
Initially, the VOI is discretized into a 3D voxel map. The reconstruction of the SPA measurements is based on the linear back-tracing of every detected photon, neutron and electron. This assumption can be justified by the fact that most secondary particles reaching the detector are produced in the peripheral areas of the cube inside the container. Particles originating from the inner areas are less likely to leave the volume of the cube due to self-shielding effects. Since no other stand-alone reconstruction approach for secondary particles exists, the presented method is the first of its kind. The proposed approach starts at the position at the innermost scintillator layer, back-traces the linear particle trajectory through the voxelized VOI and marks every voxel crossed. To assess the material characteristics per voxel, primarily the density and atomic number, the number of crossings per voxel is used as the representation value and referred to as the density score $s_{voxel}$. As the container and detector material also produce secondary particles and therefore act as a background source, the voxel map of an empty container reconstructed with 1~billion air shower events is subtracted for every scene. 

In total, twelve voxel maps are measured for each scene, which correspond to specific detector and particle measurements defined in Table~\ref{table:Measure} and denoted $M1$-$M10$. Each measurement matches the combination of a given type of secondary particle (photon, neutron, electron), its energy and the different detectors (upper detector, sidewise detectors, lower detector). This allows a discrimination between secondary particles produced and air shower particles, as well as secondary particles absorbed in the objects located inside the VOI, which is mainly achieved by separating particles with low energy from particles with high energy.

Each measurements contains a unique set of information about the object properties. For instance, low-density objects emit and absorb a significantly lower amount of neutrons than high-density materials, while the detection of low-energy secondary photons is related to the presence of a low-density material due to the increased probability of self-shielding for high-density objects. Furthermore, the absorption of air shower particles can only be measured with the lower detector, while the production of secondary particles is mainly prominent in the upper and sidewise detectors. Further information on this scheme can be found in the previous work by the authors~\cite{bib31,bib32,bib33}.

To discriminate between the materials, a unique combination of the twelve measurements is used to utilize the different kinematics of the secondary particles for each material. During this combination the density scores are summed up voxel by voxel over all selected maps. The selection of voxel maps for each combination is manually chosen based on the statistical significance of each measurement in order to avoid the injection of background noise. A further reduction of background noise is achieved by applying a minimum density score threshold $t_{min}$ for each voxel map with respect to the maximum density score $s_{max}$ in each voxel map during the combination: 

\begin{equation}
    s_{voxel} > s_{max} \cdot t_{min}
    \label{equation:Treshold}
\end{equation}

The minimum density score threshold $t_{min}$ is chosen manually for each measurement to enhance the reconstructed object over the background as much as possible while keeping the reconstructed shape and size of the object as close as possible to the original cube. The application of such a threshold is necessary as the back-tracing approach leads to a large level of noise around the reconstructed object in each voxel map. In order to remove this noise related background, a voxel is only considered for the final combined voxel map if it fulfills the threshold requirement in all of the measurements selected for the combination. The dependency of the threshold on the maximum density score was chosen to allow for a material dependent tuning of this parameter in order to improve the material discrimination. 

The set of measurements and corresponding density score thresholds used to reconstruct the cube made out of water, aluminum, vanadium, copper and lead are given in Tables~\ref{table:Water},~\ref{table:Alu},~\ref{table:Vanad},~\ref{table:Copper} and~\ref{table:Lead} respectively. The density score thresholds slowly increase for all measurements with increasing density as the reconstructed cube becomes more significant over the background due to a higher production rate of secondary particles. Neutrons in the upper and sidewise detectors are only relevant for the lead cube due to the low probability of induced fission for low and medium density materials. For all materials except lead, $M1$ and $M2.1$, as well as $M4$ and $M5.1$ are merged into one measurement to enhance the statistical significance of this voxel map during the combination procedure. 

\begin{table}[h!]
    \centering
	\caption{Definition of the measurements split by the particle type, its energy and the detector position. $M2.1$ and $M2.2$, as well as $M5.1$ and $M5.2$ are mutually exclusive.}%
	\begin{ruledtabular}
	\begin{tabular}{lccc}
		& \textbf{Photons}	& \textbf{Neutrons} & \textbf{Electrons}\\
		\hline
		\textbf{Upper detector:} & & & \\ 
		\textbf{~~~~~Production} & $M1$\footnote{ $>$~400~keV}, $M2.1$\footnote{ $<$~400~keV} & $M3$ & --\\ 
		\textbf{~~~~~Absorption} & $M2.2^{\text{b}}$ & --   & --\\ 
		\hline
		\textbf{Sidewise detector:} & & & \\ 
		\textbf{~~~~~Production} & $M4^{\text{a}}$, $M5.1^{\text{b}}$ & $M6$ & --\\ 
		\textbf{~~~~~Absorption} & $M5.2^{\text{b}}$ & --   & --\\ 
		\hline
		\textbf{Lower detector:} & & & \\ 
	    \textbf{~~~~~Production} & -- & $M8$\footnote{ $<$~3~MeV} & --\\
		\textbf{~~~~~Absorption} & $M7$	& $M9$\footnote{ $>$~3~MeV} & $M10$\\
	\end{tabular}
	\end{ruledtabular}
	\label{table:Measure}
\end{table}

\begin{table}[h!]
	\centering
	\caption{Set of measurements and density score thresholds used in the reconstruction of the water cube.}%
	\begin{ruledtabular}
	\begin{tabular}{lccc}
		& \textbf{Photons}	& \textbf{Neutrons} & \textbf{Electrons}\\
		\hline
		\textbf{Upper detector:} & & & \\
		\textbf{~~~~~Production} & $20~\%$ & -- & --\\ 
		\textbf{~~~~~Absorption} & -- & -- & --\\ 
		\hline
		\textbf{Sidewise detector:} & & & \\
		\textbf{~~~~~Production} & $30~\%$ & -- & --\\ 
		\textbf{~~~~~Absorption} & -- & -- & --\\ 
		\hline
		\textbf{Lower detector:} & & & \\
		\textbf{~~~~~Production} & -- & -- & --\\
		\textbf{~~~~~Absorption} & $30~\%$ & $40~\%$ & $40~\%$\\
	\end{tabular}
	\end{ruledtabular}
	\label{table:Water}
\end{table}

\begin{table}[h!]
	\centering
	\caption{Set of measurements and density score thresholds used in the reconstruction of the aluminum cube.}%
	\begin{ruledtabular}
	\begin{tabular}{lccc}
		& \textbf{Photons}	& \textbf{Neutrons} & \textbf{Electrons}\\
		\hline
		\textbf{Upper detector:} & & & \\
		\textbf{~~~~~Production} & $20~\%$ & -- & --\\ 
		\textbf{~~~~~Absorption} & -- & -- & --\\ 
		\hline
		\textbf{Sidewise detector:} & & & \\
		\textbf{~~~~~Production} & $30~\%$ & -- & --\\ 
		\textbf{~~~~~Absorption} & -- & -- & --\\ 
		\hline
		\textbf{Lower detector:} & & & \\
		\textbf{~~~~~Production} & -- & -- & --\\
		\textbf{~~~~~Absorption} & $30~\%$ & $40~\%$ & $40~\%$\\
	\end{tabular}
	\end{ruledtabular}
	\label{table:Alu}
\end{table}

\begin{table}[h!]
	\centering
	\caption{Set of measurements and density score thresholds used in the reconstruction of the vanadium cube.}%
	\begin{ruledtabular}
	\begin{tabular}{lccc}
		& \textbf{Photons}	& \textbf{Neutrons} & \textbf{Electrons}\\
		\hline
		\textbf{Upper detector:} & & & \\
		\textbf{~~~~~Production} & $25~\%$ & -- & --\\ 
		\textbf{~~~~~Absorption} & -- & -- & --\\ 
		\hline
		\textbf{Sidewise detector:} & & & \\
		\textbf{~~~~~Production} & $30~\%$ & -- & --\\ 
		\textbf{~~~~~Absorption} & -- & -- & --\\ 
		\hline
		\textbf{Lower detector:} & & & \\
		\textbf{~~~~~Production} & -- & -- & --\\
		\textbf{~~~~~Absorption} & $35~\%$ & $40~\%$ & $45~\%$\\
	\end{tabular}
	\end{ruledtabular}
	\label{table:Vanad}
\end{table}

\begin{table}[h!]
	\centering
	\caption{Set of measurements and density score thresholds used in the reconstruction of the copper cube.}%
	\begin{ruledtabular}
	\begin{tabular}{lccc}
		& \textbf{Photons}	& \textbf{Neutrons} & \textbf{Electrons}\\
		\hline
		\textbf{Upper detector:} & & & \\
		\textbf{~~~~~Production} & $30~\%$ & -- & --\\ 
		\textbf{~~~~~Absorption} & -- & -- & --\\ 
		\hline
		\textbf{Sidewise detector:} & & & \\
		\textbf{~~~~~Production} & $30~\%$ & -- & --\\ 
		\textbf{~~~~~Absorption} & -- & -- & --\\ 
		\hline
		\textbf{Lower detector:} & & & \\
		\textbf{~~~~~Production} & -- & -- & --\\
		\textbf{~~~~~Absorption} & $35~\%$ & $40~\%$ & $45~\%$\\
	\end{tabular}
	\end{ruledtabular}
	\label{table:Copper}
\end{table}

\begin{table}[h!]
	\centering
	\caption{Set of measurements and density score thresholds used in the reconstruction of the lead cube.}%
	\begin{ruledtabular}
	\begin{tabular}{lccc}
		& \textbf{Photons}	& \textbf{Neutrons} & \textbf{Electrons}\\
		\hline
		\textbf{Upper detector:} & & & \\
		\textbf{~~~~~Production} & $15~\%$\footnote{ $>$~400~keV}, --\footnote{ $<$~400~keV} & $15~\%$ & --\\ 
		\textbf{~~~~~Absorption} & -- & -- & --\\ 
		\hline
		\textbf{Sidewise detector:} & & & \\
		\textbf{~~~~~Production} & $25~\%^{\text{a}}$, --$^{\text{b}}$ & $25~\%$ & --\\ 
		\textbf{~~~~~Absorption} & -- & -- & --\\ 
		\hline
		\textbf{Lower detector:} & & & \\
		\textbf{~~~~~Production} & -- & -- & --\\
		\textbf{~~~~~Absorption} & $40~\%$ & $40~\%$ & $50~\%$\\
	\end{tabular}
	\end{ruledtabular}
	\label{table:Lead}
\end{table}

\subsection{\label{sec:Reco_MST}Muon scattering tomography (MST)}
Similar to the SPA, the VOI is also voxelized for the MST-based reconstruction. Two common MST algorithms are used in this work: PoCA~\cite{bib36} and ASR~\cite{bib37}. The Point-of-Closest-Approach (PoCA) method simplifies the multiple Coulomb scattering processes a muon encounters on its traversing of the examined volume by assuming that the scattering of the muon only happens at one point in the VOI. This so-called PoCA point is defined as the position in 3D space, where either the incoming and outgoing trajectory of the muon cross or where they are the closest to one another. The angle between the two trajectories is defined as the deflection of the muon and is assumed to reflect the material properties, mainly the density and atomic number~\cite{bib13,bib14,bib21}. For all measured muons, this deflection angle gets assigned to the voxel corresponding to each PoCA point and for each voxel the median over all deflection angles assigned to this voxel gets referred to as the density score $s_{voxel}$.

The method of Angle Statistics Reconstruction (ASR) is an extension to the PoCA algorithm. It also uses the angle between the incoming and outgoing muon path, but instead of assigning the scattering angle only to a single voxel, it allows to assign the deflection angle to multiple voxels in the vicinity of the two trajectories. Every voxel, which is within the minimum distance $d_{th}$ to both the incoming and outgoing path, will be populated with the measured deflection angle. In this work, $d_{th}$ is set to half of the voxel length, which was chosen manually by evaluating the significance of the reconstructed object over the background noise. Similarly to the PoCA method, the median over all deflection angles assigned to a given voxel gets referred to as the voxel density score.

\subsection{\label{sec:Reco_Combo}Muon and secondary particle combination (MST+SPA)}
To ensure a consistent comparison of the results obtained through SPA and MST, as well as the MST+SPA combination, the cubic voxel size is set to 1~dm\textsuperscript{3} for both voxel maps and the combination approach is carried out in a multi-step procedure as follows:
\begin{enumerate}
	\item {\textbf{Normalizing:} At first, the voxel maps obtained through the SPA and MST are separately normalized with respect to their maximum density score. This step ensures a common density score base for the results from SPA and MST.}
	\item {\textbf{Sharpening:} The density score in each voxel for the SPA and MST voxel maps gets reweighted by a Gaussian probability density function (PDF). The mean of the Gaussian PDF is set to the maximum normalized density score, with its FWHM reached at half of the maximum normalized density score. The sharpening is necessary to improve the significance of the reconstructed object over the background noise and to counteract the smearing of the object introduced in the smoothing step.}
	\item {\textbf{Smoothing:} A 3\texttimes3\texttimes3 convolution kernel is applied to the normalized and sharpened SPA and MST voxel maps. The kernel is configured such that the center voxel is assigned a weight of 100\%, voxels sharing a face with the center receive a weight of 50\%, those sharing an edge receive a weight of 35\% and the corner voxels are weighted by 29\%. These weights are derived from the Euclidean distance from the kernel center. The smoothing operation ensures that statistical fluctuations in the results from SPA and MST are reduced facilitating a fair and consistent comparison and combination.}
	\item {\textbf{Filtering:} The reconstructed object in the normalized, sharpened and smoothed SPA and MST voxel map is segmented from the background by applying a minimum relative density score threshold manually tuned for each material and separately for each reconstruction method (PoCA, ASR, SPA) similar to the threshold introduced in the SPA reconstruction method. The thresholds are listed in Table~\ref{table:Filter} and their values are based on the ability to reconstruct the size and shape of the cube as close as possible to the original object.}
	\item {\textbf{Aligning:} The volumetric center of the reconstructed object is aligned to the center position of the ground truth object. In both results, SPA and MST, the final reconstructed object is shifted compared to the ground truth position due to different biases in the reconstruction method~\cite{bib33}. Hence, this shift has to be eliminated for a consistent combination.}
	\item {\textbf{Merging:} Finally, the segmented voxel maps from the SPA and MST reconstructions are merged via a voxel-wise summation of their density scores. This summation is performed only for those voxels that satisfy the filtering thresholds established in the previous step. The final resulting voxel map is again normalized with respect to its maximum density score.}
\end{enumerate}

\begin{table}[h!]
    \centering
	\caption{Material dependent minimum relative density score thresholds for PoCA, ASR and SPA reconstruction for water, aluminum, vanadium, copper and lead.}%
	\begin{ruledtabular}
	\begin{tabular}{lccc}
		& \textbf{PoCA}	& \textbf{ASR} & \textbf{SPA}\\
		\hline
		\textbf{Water} & 50\% & 60\% & 50\% \\ 
		\textbf{Aluminum} & 30\% & 40\% & 50\% \\ 
		\textbf{Vanadium} & 20\% & 40\% & 50\% \\ 
		\textbf{Copper} & 20\% & 40\% & 50\% \\ 
		\textbf{Lead} & 20\% & 40\% & 50\% \\ 
	\end{tabular}
	\end{ruledtabular}
	\label{table:Filter}
\end{table}

To assess the impact of the combination, the steps 1-5 are also executed for the single SPA and MST results for a consistent comparison. The final evaluation is based on a set of geometrical performance metrics, which are based on the reconstructed and clustered object and are as follows:

\begin{itemize}
	\item {Mean density score ($s_{mean}$): Average voxel density score over all voxels associated to the reconstructed object. It serves as an indicator of the overall density distribution fidelity relative to the background.}
	\item {Volume ($v_{reco}$): Count of all voxels associated to the reconstructed object. This metric is used as an identifier for geometric artifacts resulting from the reconstruction.}
	\item {Maximum side lengths ($l_{x}$, $l_{y}$, $l_{z}$): Difference between the maximum and minimum x-, y- or z-positions of the object. Similar to $v_{reco}$, the side lengths capture the spatial relationship and accuracy of the object.}
	\item {Chamfer distance ($d_{c}$): Geometrical difference between the reconstructed and ground truth object~\cite{bib32,bib38,bib39}. When comparing objects of similar size, a smaller $d_{c}$ indicates a closer match in shape between the objects. If however the compared objects vary significantly in size, the Chamfer distance will not only reflect the shape differences, but also the variation in size. Hence, this metric is evaluated in the context of relative comparisons rather than as an absolute measure.}
\end{itemize}

These metrics provide a robust framework for the evaluation of the multiple aspects of the reconstruction algorithm between stand-alone methodologies (SPA or MST) and the integrated MST+SPA approach.

\section{\label{section:Res}Results}
\begin{figure}[h!]
	\centering
	\includegraphics[width=8.6cm]{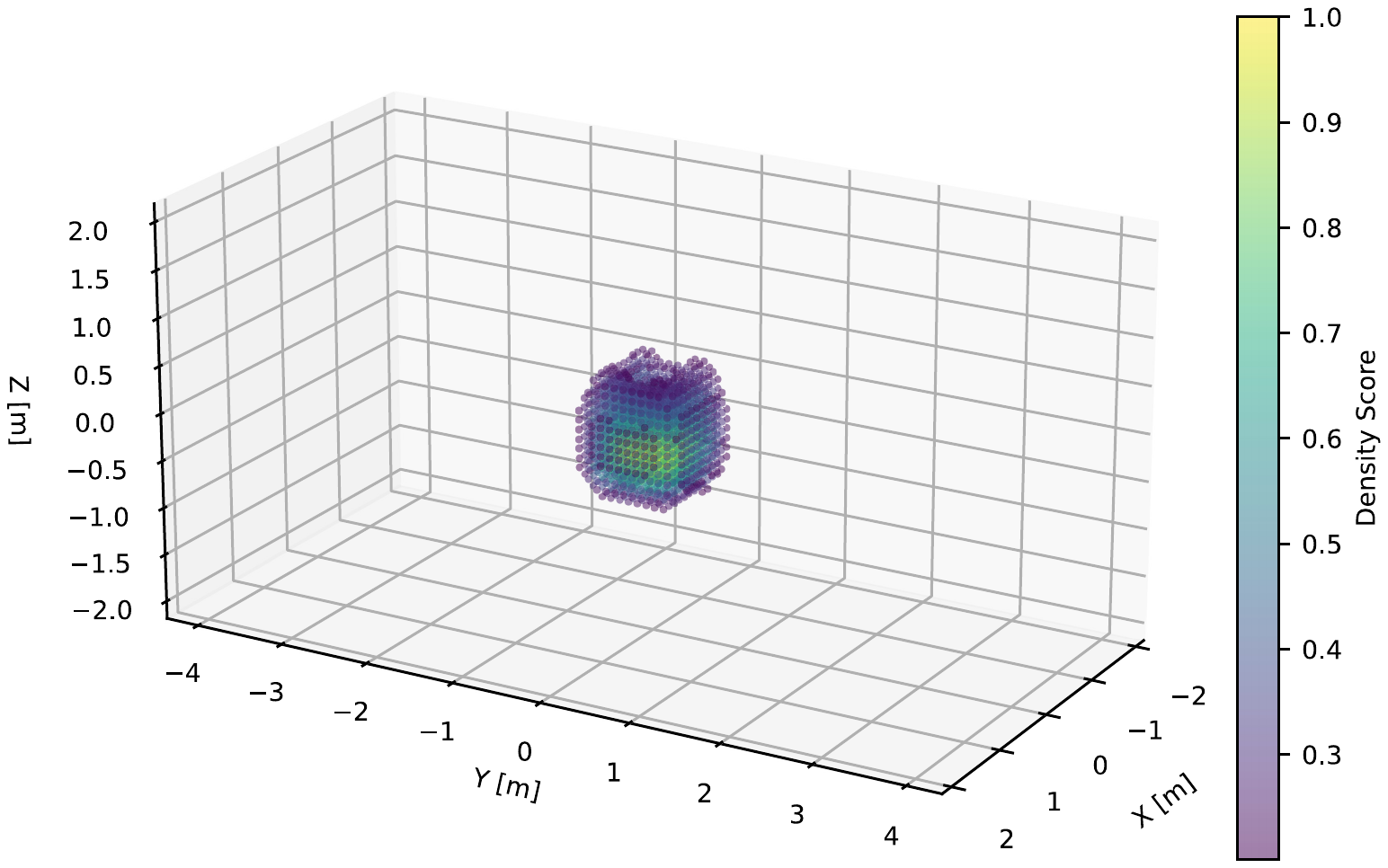}
	\caption{Reconstruction of the vanadium cube inside the container using the PoCA method.}
	\label{figure:Vanad_PoCA}
\end{figure}

\begin{figure}[h!]
	\centering
	\includegraphics[width=8.6cm]{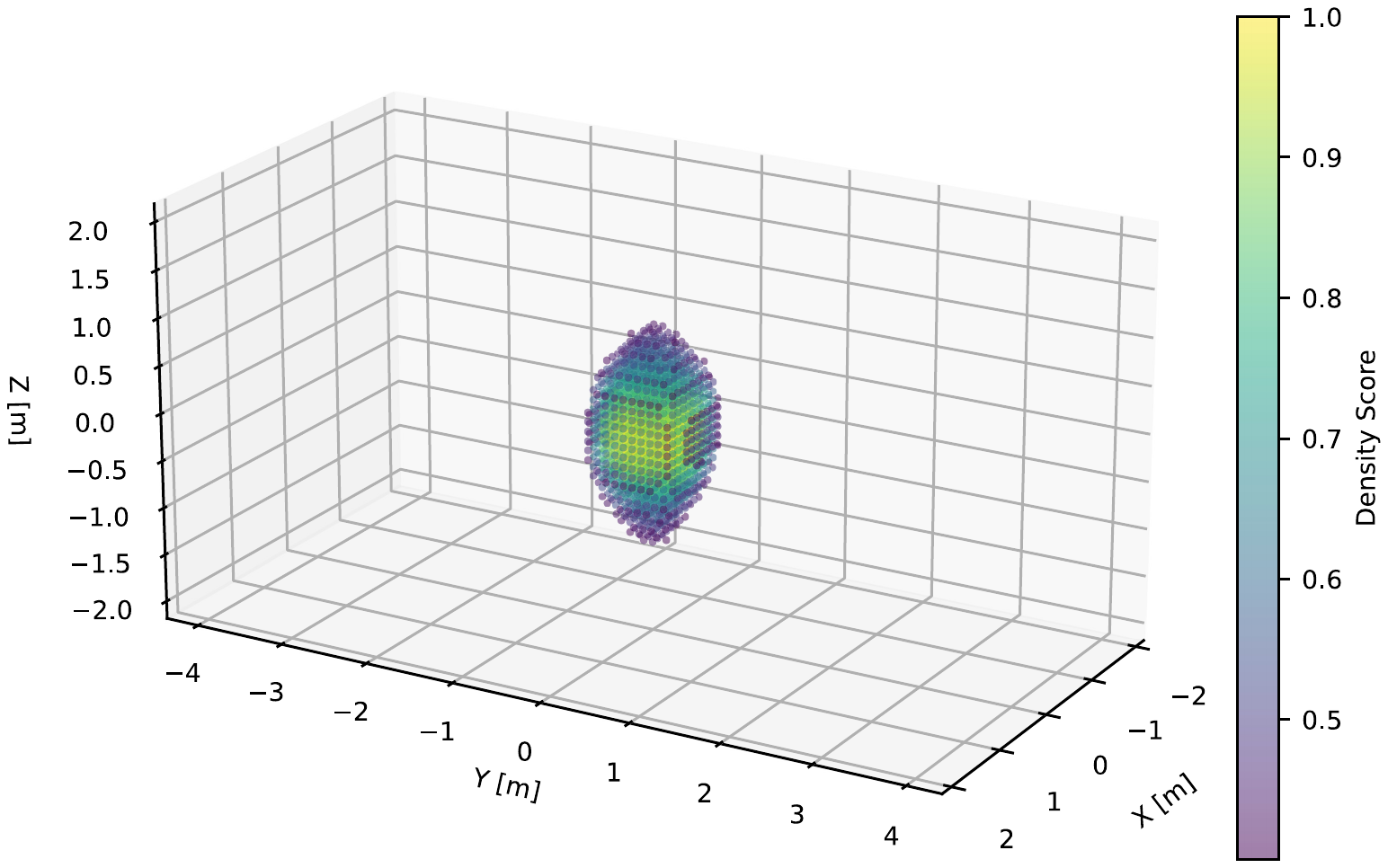}
	\caption{Reconstruction of the vanadium cube inside the container using the ASR method.}
	\label{figure:Vanad_ASR}
\end{figure}

\begin{figure}[h!]
	\centering
	\includegraphics[width=8.6cm]{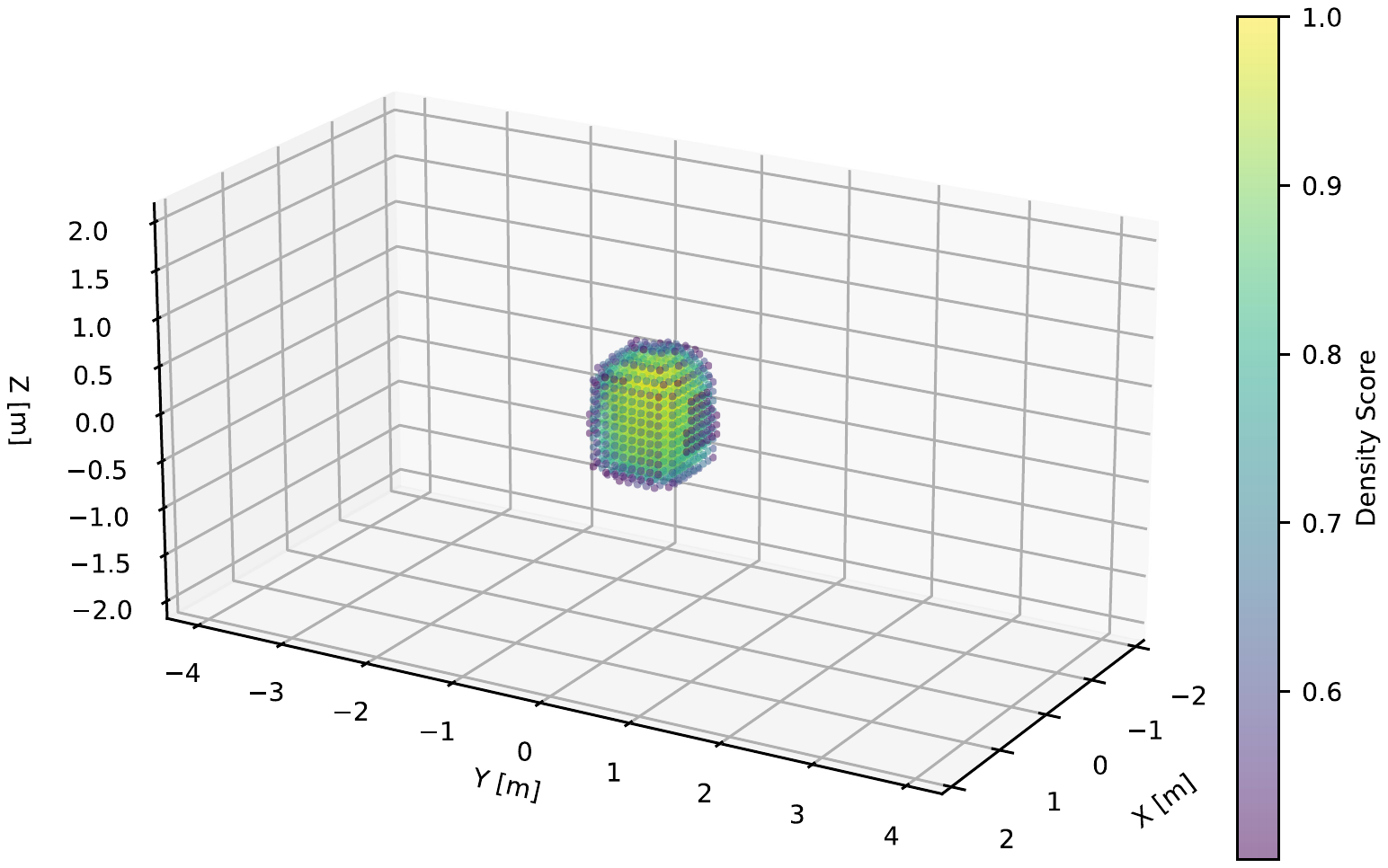}
	\caption{Reconstruction of the vanadium cube inside the container using the SPA method.}
	\label{figure:Vanad_SPA}
\end{figure}

\begin{figure}[h!]
	\centering
	\includegraphics[width=8.6cm]{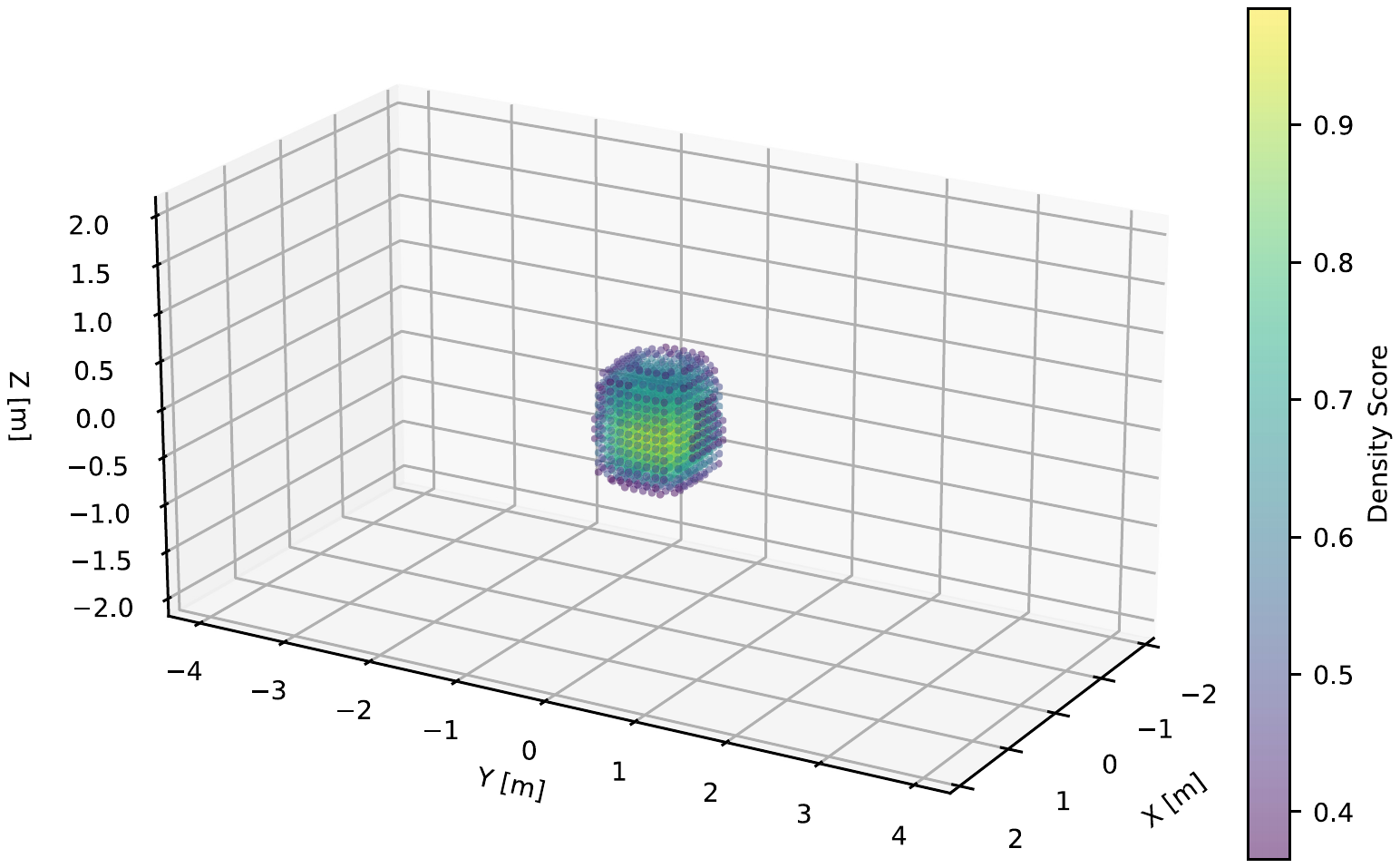}
	\caption{Reconstruction of the vanadium cube inside the container using the PoCA+SPA combination.}
	\label{figure:Vanad_PoCA-SPA}
\end{figure}

\begin{figure}[h!]
	\centering
	\includegraphics[width=8.6cm]{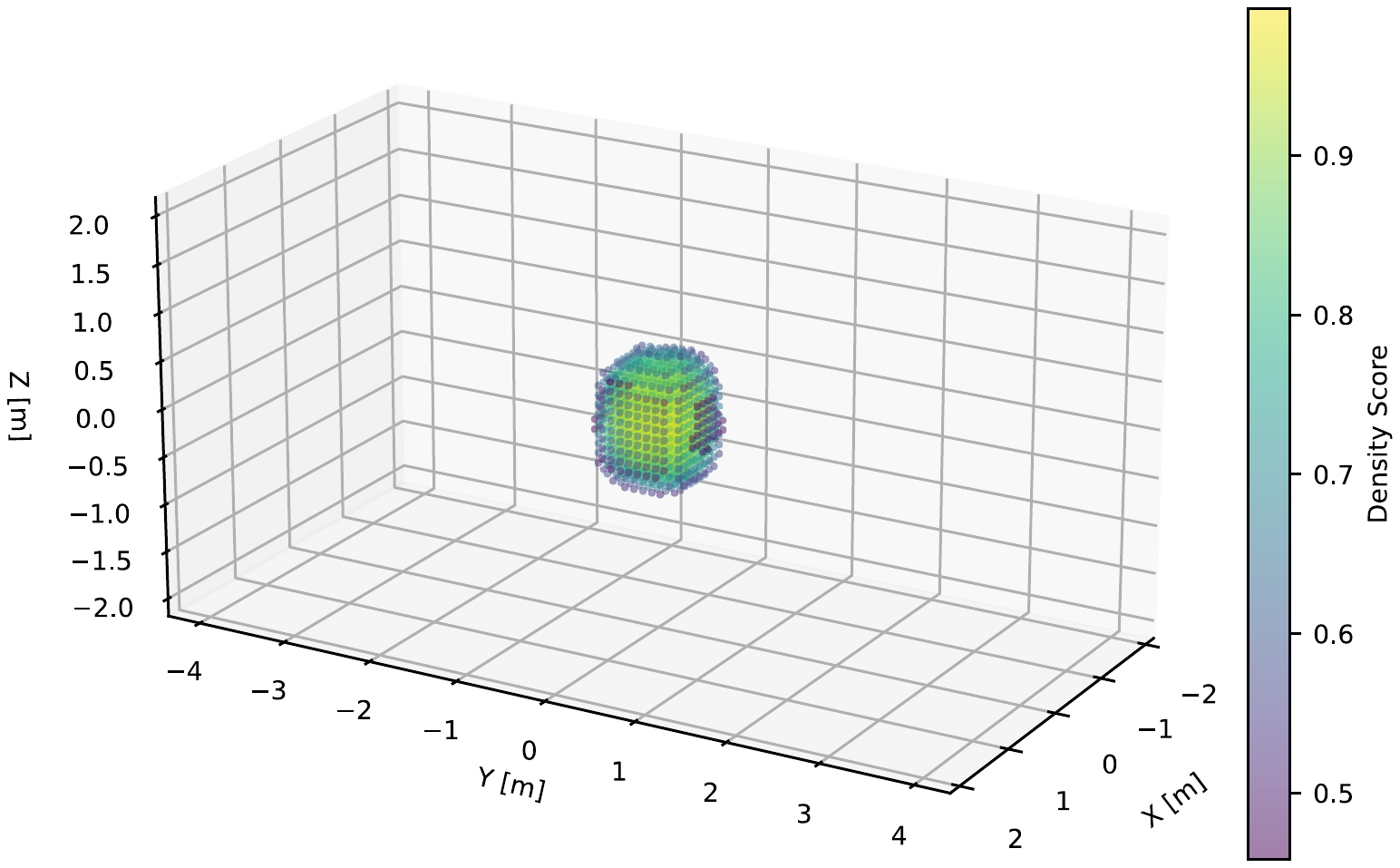}
	\caption{Reconstruction of the vanadium cube inside the container using the ASR+SPA combination.}
	\label{figure:Vanad_ASR-SPA}
\end{figure}

The results of the single reconstructions, as well as the MST+SPA combinations for the five different scenes are shown in terms of the performance metrics in Tables~\ref{table:Res_Water}--\ref{table:Res_Lead}. For comparison, the volume of the ground truth object is 1.21~m\textsuperscript{3}, while the side lengths are 1.1~m ($l_{x}$ and $l_{z}$) and 1.0~m ($l_{y}$). The difference with respect to the simulated cube of 1.0~m~\texttimes~1.0~m~\texttimes~1.0~m~ occurs due to the necessary voxelization of the object to be comparable to the reconstructed voxel map. This effect can be seen for the cases, where the edges of the simulated cube are not consistent with the edges of the voxel grid, letting the voxelized version of the object appear bigger than its original. For visualization, the different voxel maps of the MST and SPA reconstruction methods, as well as the MST+SPA combinations of the vanadium cube are shown in Figures~\ref{figure:Vanad_PoCA}--\ref{figure:Vanad_ASR-SPA}.

\begin{table}[h!]
    \centering
	\caption{Results of the reconstruction of the water cube using PoCA, ASR, SPA, PoCA+SPA and ASR+SPA.}%
	\begin{ruledtabular}
	\begin{tabular}{lccccc}
		& \textbf{PoCA}	& \textbf{ASR} & \textbf{SPA} & \textbf{PoCA+SPA} & \textbf{ASR+SPA}\\
		\hline
		\textbf{$s_{mean}$} & 0.69 & 0.81 & 0.78 & 0.75 & 0.83 \\ 
		\textbf{$v_{reco}$} & 1.19~m\textsuperscript{3} & 1.89~m\textsuperscript{3} & 1.18~m\textsuperscript{3} & 1.09~m\textsuperscript{3} & 1.18~m\textsuperscript{3} \\ 
		\textbf{$l_{x}$} & 1.1~m & 1.1~m & 1.1~m & 1.1~m & 1.1~m \\ 
		\textbf{$l_{y}$} & 1.2~m & 1.0~m & 1.0~m & 1.0~m & 1.0~m \\ 
		\textbf{$l_{z}$} & 1.4~m & 2.5~m & 1.4~m & 1.4~m & 1.4~m \\ 
		\textbf{$d_{c}$} & 4.2 & 44.3 & 4.1 & 3.8 & 4.1 \\ 
	\end{tabular}
	\end{ruledtabular}
	\label{table:Res_Water}
\end{table}

\begin{table}[h!]
    \centering
	\caption{Results of the reconstruction of the aluminum cube using PoCA, ASR, SPA, PoCA+SPA and ASR+SPA.}%
	\begin{ruledtabular}
	\begin{tabular}{lccccc}
		& \textbf{PoCA}	& \textbf{ASR} & \textbf{SPA} & \textbf{PoCA+SPA} & \textbf{ASR+SPA}\\
		\hline
		\textbf{$s_{mean}$} & 0.56 & 0.69 & 0.80 & 0.71 & 0.80 \\ 
		\textbf{$v_{reco}$} & 1.55~m\textsuperscript{3} & 2.07~m\textsuperscript{3} & 1.23~m\textsuperscript{3} & 1.22~m\textsuperscript{3} & 1.23~m\textsuperscript{3} \\ 
		\textbf{$l_{x}$} & 1.3~m & 1.1~m & 1.1~m & 1.1~m & 1.1~m \\ 
		\textbf{$l_{y}$} & 1.2~m & 1.2~m & 1.0~m & 1.0~m & 1.0~m \\ 
		\textbf{$l_{z}$} & 1.5~m & 2.6~m & 1.5~m & 1.3~m & 1.5~m \\ 
		\textbf{$d_{c}$} & 2.4 & 52.1 & 1.5 & 1.3 & 1.5 \\ 
	\end{tabular}
	\end{ruledtabular}
	\label{table:Res_Alu}
\end{table}

\begin{table}[h!]
    \centering
	\caption{Results of the reconstruction of the vanadium cube using PoCA, ASR, SPA, PoCA+SPA and ASR+SPA.}%
	\begin{ruledtabular}
	\begin{tabular}{lccccc}
		& \textbf{PoCA}	& \textbf{ASR} & \textbf{SPA} & \textbf{PoCA+SPA} & \textbf{ASR+SPA}\\
		\hline
		\textbf{$s_{mean}$} & 0.46 & 0.71 & 0.81 & 0.67 & 0.79 \\ 
		\textbf{$v_{reco}$} & 1.65~m\textsuperscript{3} & 1.75~m\textsuperscript{3} & 1.35~m\textsuperscript{3} & 1.30~m\textsuperscript{3} & 1.32~m\textsuperscript{3} \\ 
		\textbf{$l_{x}$} & 1.3~m & 1.1~m & 1.1~m & 1.1~m & 1.1~m \\ 
		\textbf{$l_{y}$} & 1.2~m & 1.2~m & 1.2~m & 1.2~m & 1.2~m \\ 
		\textbf{$l_{z}$} & 1.4~m & 2.3~m & 1.4~m & 1.3~m & 1.4~m \\ 
		\textbf{$d_{c}$} & 5.8 & 21.8 & 4.7 & 1.2 & 4.6 \\ 
	\end{tabular}
	\end{ruledtabular}
	\label{table:Res_Vanad}
\end{table}

\begin{table}[h!]
    \centering
	\caption{Results of the reconstruction of the copper cube using PoCA, ASR, SPA, PoCA+SPA and ASR+SPA.}%
	\begin{ruledtabular}
	\begin{tabular}{lccccc}
		& \textbf{PoCA}	& \textbf{ASR} & \textbf{SPA} & \textbf{PoCA+SPA} & \textbf{ASR+SPA}\\
		\hline
		\textbf{$s_{mean}$} & 0.45 & 0.71 & 0.82 & 0.65 & 0.79 \\ 
		\textbf{$v_{reco}$} & 1.52~m\textsuperscript{3} & 1.67~m\textsuperscript{3} & 1.41~m\textsuperscript{3} & 1.30~m\textsuperscript{3} & 1.34~m\textsuperscript{3} \\ 
		\textbf{$l_{x}$} & 1.3~m & 1.1~m & 1.1~m & 1.1~m & 1.1~m \\ 
		\textbf{$l_{y}$} & 1.2~m & 1.2~m & 1.2~m & 1.2~m & 1.2~m \\ 
		\textbf{$l_{z}$} & 1.4~m & 2.2~m & 1.4~m & 1.3~m & 1.4~m \\ 
		\textbf{$d_{c}$} & 5.0 & 21.1 & 5.0 & 1.2 & 4.8 \\ 
	\end{tabular}
	\end{ruledtabular}
	\label{table:Res_Copper}
\end{table}

\begin{table}[h!]
    \centering
	\caption{Results of the reconstruction of the lead cube using PoCA, ASR, SPA, PoCA+SPA and ASR+SPA.}%
	\begin{ruledtabular}
	\begin{tabular}{lccccc}
		& \textbf{PoCA}	& \textbf{ASR} & \textbf{SPA} & \textbf{PoCA+SPA} & \textbf{ASR+SPA}\\
		\hline
		\textbf{$s_{mean}$} & 0.44 & 0.70 & 0.81 & 0.64 & 0.78 \\ 
		\textbf{$v_{reco}$} & 1.35~m\textsuperscript{3} & 1.51~m\textsuperscript{3} & 1.43~m\textsuperscript{3} & 1.24~m\textsuperscript{3} & 1.32~m\textsuperscript{3} \\ 
		\textbf{$l_{x}$} & 1.1~m & 1.1~m & 1.1~m & 1.1~m & 1.1~m \\ 
		\textbf{$l_{y}$} & 1.2~m & 1.2~m & 1.2~m & 1.2~m & 1.2~m \\ 
		\textbf{$l_{z}$} & 1.3~m & 1.9~m & 1.4~m & 1.3~m & 1.4~m \\ 
		\textbf{$d_{c}$} & 1.1 & 8.2 & 5.1 & 1.1 & 4.8 \\ 
	\end{tabular}
	\end{ruledtabular}
	\label{table:Res_Lead}
\end{table}

The results of the reconstruction of the water cube in Table~\ref{table:Res_Water} show overall a noticeable improvement for most of the metrics compared to the single MST and SPA results. The volumetric size of the reconstructed object is below the ground-truth value of 1.21~m\textsuperscript{3}, except for the ASR result, which overestimates the volume at 1.89~m\textsuperscript{3}. This is due to the significantly elongated length in the z-direction, which is a known artifact of the ASR algorithm and can be seen for all materials studied in this work. This also leads to a significantly higher Chamfer distance of 44.3 for ASR compared to the other algorithms. It can also be seen that the combined results are comparable for PoCA+SPA and ASR+SPA, which suggests that the artifacts in the ASR reconstruction can be effectively reduced by the combination with the SPA result.

For the scene with the aluminum cube, a similar noticeable improvement for most of the metrics in Table~\ref{table:Res_Alu} can be seen similar to the water cube scenario. However, the volumetric size of the reconstructed object is now matching the truth value for the combined measurements and SPA, but is higher for PoCA and ASR. While the average density score of the combined results is lower than for the water case, suggesting a less even distribution of the reconstructed density in the clustered object, $v_{reco}$ and $d_{c}$ show more similarity to the ground truth object.

The reconstruction of the vanadium cube can also be improved with the combined approach in comparison to the single results as shown in Table~\ref{table:Res_Vanad}. For this scene, the volumetric size of the reconstructed object is overall higher than the truth value for all measurements. In addition, it becomes evident that compared to PoCA and ASR the SPA method shows the best performance in reconstructing a high density score, which is related to an even density score distribution of the cube, while PoCA is performing the poorest in this metric in general. This is also expected as the PoCA algorithm is known to show uneven density reconstruction due to its very simplified approach of assigning the scattering only to one voxel. PoCA and SPA show also clearly the highest similarity for the geometrical metrics compared to the ground truth object, while ASR is failing in these cases due to the already discussed artifact along the z-axis. 

The combined results in Table~\ref{table:Res_Copper} show an improvement in most of the metrics even for the copper scene. The metrics are similar to the vanadium setup suggesting that the reconstruction methods are reaching their optimal working conditions for materials with medium densities. This can also be seen in the reconstructed shape along the z-direction for the ASR results, which is the worst for lower density material cubes.

Also for the highest density cube made out of lead, the combination results are showing an overall performance improvement for the majority of metrics compared to the single SPA, PoCA and ASR measurements as shown in Table~\ref{table:Res_Lead}. However, it is noticeable that with increasing density $s_{mean}$ drops continuously for the single, as well as for the combined results. This is likely the effect of various simplification assumptions, which are part of all reconstruction algorithms, leading to less continuous density distributions inside the reconstructed cube.

\section{\label{section:Con}Conclusion}
This work presented the first comparison and combination of stand-alone muon and secondary particle results in the field of cosmic-ray tomography by analyzing the contents of a shipping container with a simplified detection and geometry setup. The simulation-based study evaluated the performance of the PoCA, ASR and SPA reconstruction methods, as well as the combinations of PoCA+SPA and ASR+SPA to reconstruct a 1~m\textsuperscript{3} cube made out of five different materials inside a shipping container. In order to ensure a consistent comparison and combination between the different methods, the study follows a multi-step procedure to segment the reconstructed cube over the background. The assessment of the performance of the different reconstruction algorithms is based on dedicated properties of the reconstructed object, such as its density score and other geometric quantities.

The results demonstrate that incorporating secondary particle information via this novel dual-channel analysis technique can serve as a valuable complementary data source. Noticeable improvements for all performance metrics over single MST and SPA results are achieved reducing known reconstruction artifacts of the utilized reconstruction algorithms. The improvements persist across a range of materials and densities, thus highlighting the applicability of the proposed method.

However, the current methodology exhibits several constraints, for instance the lack of a generalization and automatization of the approach. Some steps currently require the manual tuning of reconstruction parameters, like the minimum density thresholds or the geometric corrections used in the combination procedure, which are all varying for the position and material of the objects studied in this work. 

Future work will address these limitations by developing a machine learning–based framework capable of automatically optimizing the necessary settings using a diverse training dataset generated via the B2G4 synthetic data pipeline~\cite{bib40}. Additionally, a generalized material classification algorithm is needed to efficiently categorize objects and assess density distributions across both single-channel and combined MST+SPA voxel maps~\cite{bib41}. Last, the simplified approach used for the detector and scene geometry will be improved by the usage of more complex objects with the B2G4 package and by experimentally measuring secondary particle detection efficiencies for state-of-the-art commercial muon detectors using radiation sources. 

In conclusion, the simulation-based studies on the potential utilization of the complementary information of secondary particles show that such an additional analysis can lead to visible benefits in the reconstruction of hidden objects in the context of cosmic-ray tomography. While the theoretical base of this approach seems to be established now, experimental studies are crucial for the next steps of this dual-channel approach towards potential applications in the future.

\begin{acknowledgments}
This research was funded by the SilentBorder project under the grant agreement ID 101021812 of the European Union’s Horizon 2020 research and innovation program. Assistance during the creation of the geometry visualization provided by Felix Sattler was greatly appreciated.
\end{acknowledgments}

\section*{Conflict of Interest Statement}
The authors have no conflicts to disclose.

\section*{Author Contributions}
\textbf{Maximilian P\'{e}rez Prada:} Conceptualization (supporting), Data Curation (lead), Formal Analysis (lead), Investigation (lead), Methodology (lead), Software (lead), Validation (equal), Visualization (lead), Writing -- Original Draft Preparation (lead). 

\textbf{\'{A}ngel Bueno Rodr\'{i}guez:} Methodology (supporting), Validation (equal), Writing -- Review and Editing (equal). 

\textbf{Maurice Stephan:} Conceptualization (supporting), Funding Acquisition (supporting), Project Administration (supporting), Resources (lead), Supervision (supporting), Writing -- Review and Editing (equal).

\textbf{Sarah Barnes:} Conceptualization (lead), Funding Acquisition (lead), Methodology (supporting), Project Administration (lead), Resources (supporting), Supervision (lead), Validation (equal), Writing -- Review and Editing (equal). 

\section*{Data Availability Statement}
The data that support the findings of this study are available from the corresponding author upon reasonable request.

\end{document}